\documentclass[5p,times,twocolumn]{elsarticle}

\usepackage[hidelinks]{hyperref}
\usepackage{amssymb}
\usepackage{booktabs,adjustbox} 
\usepackage{siunitx}
\usepackage[UKenglish]{isodate}
\usepackage{graphicx}
\usepackage[fleqn]{mathtools} 
\usepackage{xcolor}
\usepackage[UKenglish]{isodate}
\cleanlookdateon
\usepackage{caption,subcaption}
\usepackage{numcompress}\biboptions{authoryear}

\emergencystretch=\maxdimen
\makeatletter
\def\ps@pprintTitle{%
   \let\@oddhead\@empty
   \let\@evenhead\@empty
   \let\@oddfoot\@empty
   \let\@evenfoot\@oddfoot
}
\makeatother

\begin{document}

\begin{frontmatter}
\title{The pear-shaped fate of an ice melting front}
\author[ouraddress]{James N. Hewett\corref{mycorrespondingauthor}}
\ead{james@hewett.nz}
\cortext[mycorrespondingauthor]{Corresponding author}
\author[ouraddress]{Mathieu Sellier}
\address[ouraddress]{Department of Mechanical Engineering, University of Canterbury, Christchurch 8140, New Zealand}

\begin{abstract}
A fluid-structure interaction problem with the melting of water around a heated horizontal circular cylinder is analysed with numerical simulations. Dynamic meshing was used for evolving the flow domain in time as the melting front extended radially outward from the cylinder; a node shuffle algorithm was used to retain mesh quality across the significant mesh deformation. We simulated one case above the density inversion point of water and one case below, yielding pear-shaped melting fronts due to thermal plumes either rising or falling from the cylinder, respectively. Results were compared with previous experimental studies and the melting front profiles matched reasonably well and melting rates were in agreement. We confirm that natural convection plays a significant role in the transport of energy as the melt zone increases, and needs to be considered for accurately modelling phase change under these conditions.
\end{abstract}
\end{frontmatter}

\section{Introduction}
Thermal energy storage plays an important role in utilising energy resources effectively because often the timing of generation and consumption of energy can vary from hours to months. For example, solar energy is only available during the day and therefore effective energy storage is required for utilising solar energy during the night. Similarly, power stations must design for peak loads whereas adequate energy storage would allow more efficient use of generators as peak times can be offset by stored energy from off peak times. An efficient method of storing thermal energy is with latent heat which provides a high storage density and requires a smaller difference between storing and releasing temperatures compared with the sensible heat storage method \citep{Farid2004}. There has also been recent studies on using phase change materials for passive cooling in buildings where latent heat is used to increase the thermal inertia of building envelopes, regularising the ambient temperature \citep{Akeiber2016}.

The moving boundary problem where the solid and liquid phase change process occurs forms the classical Stefan problem \citep{Stefan1891}; named after the Slovene physicist Jo$\breve{\text{z}}$ef Stefan (1835-1893). This moving boundary is a function of both time and space and is unknown a priori; creating a coupled fluid structure interaction problem to model.

Heat transfer by conduction dominates as the mechanism responsible for melting in the initial stages, when a thin layer of water is present. However, natural convection (due to the temperature dependent water density) plays an important role as the melting front advances outward from the heat source and the volume of the flow domain increases \citep{White1977,Sparrow1978,Bathelt1979}. The flow induced from these buoyancy effects creates temperature fields which lead to a \textit{pear-shaped} solid-liquid interface. Water has a density inversion near $\SI{4.0}{\celsius}$ which influences the location of enhanced melting (either the warmer water rises or falls around the cylinder). An inverted pear-shaped interface was found for cylinder temperatures below $\SI{8.0}{\celsius}$ \citep{Herrmann1984} and near concentric interface evolution at $\SI{8.0}{\celsius}$. We simulate one case below and another above this critical cylinder temperature to explore both scenarios due to the density inversion of water.

Experiments have previously been undertaken on the melting of phase change materials around horizontal cylindrical heat sources of: \textit{n}-paraffins (\textit{n}-heptadecane and \textit{n}-octadecane) \citep{Bathelt1980} where no density inversion exists; and water \citep{White1986} at temperatures around the density inversion point. We simulate water as the phase change material and quantitatively compare our results with the latter set of experiments.

Previous simulation approaches include numerical mapping techniques where the transformed domain morphs over time \citep{Rieger1982,Ho1986}, and another by using the latent heat content which varies between zero (solid) and 1 (liquid) \citep{Darzi2012}. The first approach involves tracking the melting interface via domain mapping and calculating the governing equations on a stationary grid. The second approach employs a single mesh where cells have a liquid fraction assigned and the interface is determined based on a fraction threshold criteria. Alternatively, we directly tracked the moving boundary by dynamically updating the mesh throughout the simulation based on the heat flux at the solid-liquid interface.

\section{Methods}

\subsection{Problem description}
The physical problem studied in this paper is of ice melting radially outward from a horizontal isothermal heated cylinder. Initially, the ice has a uniform temperature at its fusion point $T_f$ and the cylinder is heated to $T_w > T_f$. We model the molten ice as a single phase with the interface boundary imitating the Stefan condition. The interface is tracked across discrete time steps where the mesh dynamically updates. Flow is solved in steady state at each time step, leading to a quasi-steady state simulation.

Considering the complexity of modelling phase change materials with dynamic boundaries, the following assumptions were made: (1) motion of water is laminar, 2-D and is incompressible; (2) thermophysical properties of water, except density, are constant across the temperature range modelled; (3) the Boussinesq approximation (density variations only feature in the buoyancy source term); (4) viscous dissipation and volume difference due to phase change are neglected; and (5) thermal equilibrium exists at the interface.

\subsubsection{Geometry and boundary conditions}
The heated cylinder, bounded with a radius of $R_w = \SI{12.7}{\milli\metre}$, is located at the centre of the computational domain and remains stationary. We selected two cylinder wall temperature values: $T_w = \SI{2.3}{\celsius}$ and $T_w = \SI{14.1}{\celsius}$, to directly compare our simulation results with existing experimental data \citep{White1986}; one below the minimum density point $T_m = \SI{4.029325}{\celsius}$ (heat source sinks) and one above $T_m$ (heat source rises).

Ice surrounds the cylinder outward to infinity and the interface between the ice and water is described with the radial distance $R_i(\theta,\tau)$ where $\theta$ is the angle measured from the base of the cylinder. The computational domain resides between the heated cylinder wall boundary and the dynamic melting interface.

A dimensionless melted ice volume ratio $V$ was defined as the ratio of molten ice to the isothermal cylinder, calculated with
\begin{equation}\label{eqn:volume}
V = \frac{A_i - A_w}{A_w}
\end{equation}
using cross-sectional areas $A$ because the depth is an arbitrary parameter for our 2-D model. The area enclosed by the interface $A_i$ was calculated by treating the nodes outlining the boundary as a simple polygon, and the constant wall area was calculated with $A_w = \pi R_w^2$.

\begin{figure}\centering
  \begin{subfigure}[b]{0.25\textwidth} \centering
    \includegraphics{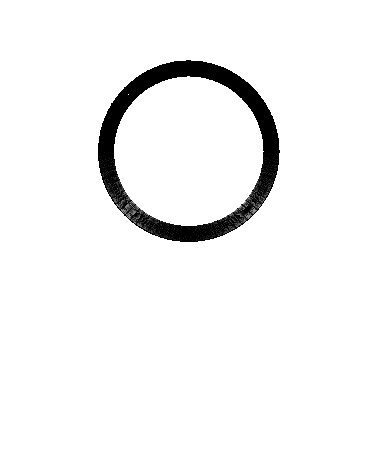}
  \end{subfigure}%
  \begin{subfigure}[b]{0.25\textwidth} \centering
    \includegraphics{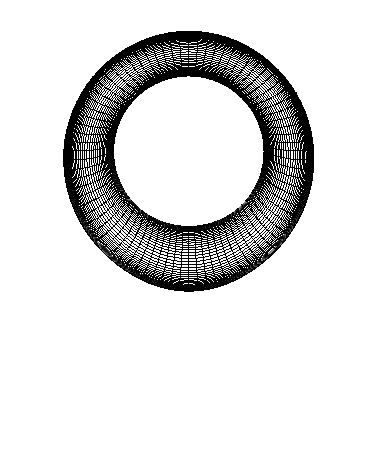}
  \end{subfigure}
  \begin{subfigure}[b]{0.25\textwidth} \centering
    \includegraphics{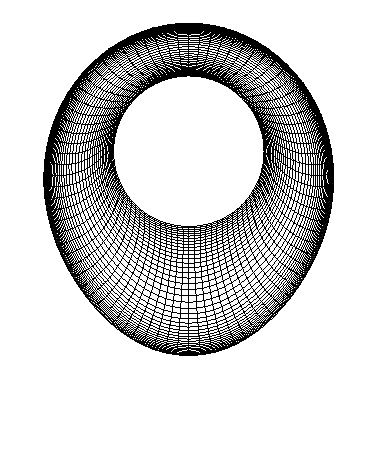}
  \end{subfigure}%
  \begin{subfigure}[b]{0.25\textwidth} \centering
    \includegraphics{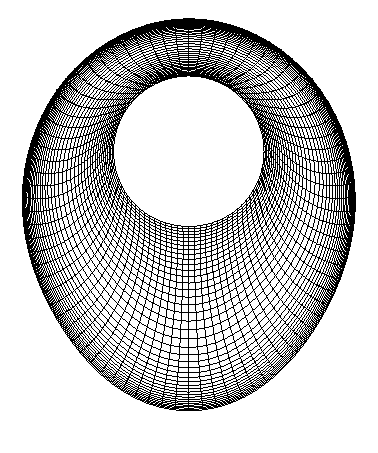}
  \end{subfigure}
  \caption{Computational mesh evolution from initial (top left) to final (bottom right) geometry for one of the simulations.\label{fig:FigureH}}
\end{figure}

A structured O-grid mesh of $40 \times 80$ (radial $\times$ circumferential) was used in all of the simulations. Uniform cell lengths were specified around the perimeter and a bias was applied in the radial direction to cluster cells near both of the boundaries; in order to resolve the thermal boundary layers. The outer boundary (solid and molten ice interface) expands and the mesh dynamically updates accordingly throughout the simulation as shown in Figure~\ref{fig:FigureH}. Computational cells within the mesh are deformed and translated without creating or destroying cells; a constant number of finite volume cells are retained throughout each simulation.

Dirichlet conditions were applied to the cylinder wall and interface boundaries with temperatures of $T_w$ and $T_i$ respectively, and no slip shear conditions were also imposed. The front and back boundaries of the O-grid domain were set as symmetry with one cell depth to enforce the 2-D assumption.

\subsubsection{Initial conditions}
The experiments \citep{White1986} started with the solid ice in contact with the cylinder such that the initial interface radius $R_0 = R_w$ (no molten ice was present). However, our simulations require a finite volume to begin with and therefore we used $R_0 = 1.2 R_w$ and offset the time appropriately with $t_0$ (as calculated based on the analytical solution of heat transfer by conduction described in Section~\ref{sec:SectionA}.

As the heat transport within the molten ice is dominated by conduction in the initial stages, we initialised the computational domain with the analytical solution to this conduction problem. Velocities were set to zero and the temperature field was initialised using
\begin{equation}\label{eqn:conduction}
T = (T_i - T_w) \frac{\ln{(r/R_w)}}{\ln{(R_i/R_w)}} + T_w
\end{equation}
where $r$ is the radial coordinate; with the substitution $R_i = R_0$.

\subsubsection{Fluid properties}
The nonlinear variation of water density $\rho$ was included in our simulations by using a $\rho$ relation in the range of $T = \SI{0}{}$ to $\SI{20}{\celsius}$ \citep{Gebhart1977}
\begin{equation}\label{eqn:density}
\rho = \rho_m (1 - C_T | T - T_m |^q)
\end{equation}
where $\rho_m = \SI{999.9720}{\kilogram/\metre\cubed}$, $C_T = \SI{9.297173e-6}{/\kelvin}$ is the temperature coefficient, $T_m = \SI{4.029325}{\celsius}$ and $q = \SI{1.894816}{}$ the temperature index.

Thermophysical properties of water were evaluated at an average temperature of $T \approx \SI{5}{\celsius}$: specific heat $c_p = \SI{4.20}{\kilo\joule/\kilogram.\kelvin}$, dynamic viscosity $\mu = \SI{1.52}{\gram/\metre.\second}$ and thermal conductivity $k = \SI{0.57}{\watt/\metre.\kelvin}$. These properties vary slightly across the temperature range simulated but have a negligible influence on the melting rate compared to the density variation.

The constant temperature of the heated cylinder $T_w$ was non-dimensionalised with the Stefan number
\begin{equation}\label{eqn:Stefan}
\text{Ste} = \frac{c_p (T_w - T_f)}{\Delta h_f}
\end{equation}
where $\Delta h_f = \SI{333.55}{\kilo\joule/\kilogram}$ is the latent heat of fusion. The time $t$ was made dimensionless with the product of the Fourier and Stefan numbers with
\begin{equation}\label{eqn:tau}
\tau = \frac{\alpha t}{R_w^2} \text{Ste}
\end{equation}
where $\alpha = k/\rho c_p$ is the thermal diffusivity.

The Rayleigh number is a measure of the intensity of natural convection within the molten ice. A density based definition \citep{White1986} was chosen to handle the non-linear density variation and density inversion feature, with
\begin{equation}\label{eqn:Rayleigh}
\text{Ra} = \frac{g R_w^3 (\rho_m - \rho_\text{film})}{\nu \alpha \rho_m}
\end{equation}
where $g = \SI{9.80665}{\metre/\second\squared}$ is the acceleration due to gravity, $\nu$ the kinematic viscosity and $\rho_\text{film}$ the density at the film temperature $T_\text{film} = (T_i + T_f)/2$. $\text{Ra}$ approaches zero at the density inversion point ($T_w = \SI{8.0}{\celsius}$ and therefore $T_\text{film} = \SI{4.0}{\celsius}$) where natural convection plays an insignificant role. Conversely, $\text{Ra}$ increases further from this point and is positive for both upright and inverted pear-shaped melting fronts.

\subsection{Numerical procedure}
Our simulations were performed using ANSYS Fluent R17.0 as the computational fluid dynamics software. Data analysis and visualisation of results were coded in MATLAB 2016b.

\subsubsection{Governing equations}
Fluent is a cell centred finite volume solver and was employed to solve the momentum, continuity and energy equations. Second order spatial discretisation methods were set for the pressure, momentum and energy equations. Pressure and velocity were coupled with the PISO scheme. Under-relaxation factors of 0.3 (pressure), 1 (density), 1 (body forces), 0.7 (momentum) and 0.7 (energy) were used.

The fluid time step specified in Fluent was $\SI{1e6}{\second}$ (quasi-steady state assumption) whereas the dynamic mesh step used for deforming the interface was $\Delta t = \SI{250}{\second}$ ($\tau = 0.006$) resulting in 120 steps for the inverted pear-shape and $\Delta t = \SI{20}{\second}$ ($\tau = 0.003$) with 200 steps for the upright pear-shape case. A maximum number of 50 iterations per time step was used as this number gave iterative convergence of the solution.

\subsubsection{Melting interface boundary}
The velocity of the melting front at the solid-liquid water interface was given by the Stefan condition \citep{Moore2017}
\begin{equation}\label{eqn:velocity}
\textbf{v}_i =  - \frac{\alpha c_p}{\Delta h_f} \frac{dT}{dn} \Big|_i \hat{\textbf{n}}
\end{equation}
where $\hat{\textbf{n}}$ is the unit normal vector to the interface and is directed toward the solid phase. The $\textbf{v}_i$ was positive for all cases because a negative temperature gradient existed at the interface boundary; yielding an outward melting front from the warm cylinder throughout the simulations.

\subsubsection{Dynamic mesh}
The dynamic mesh model in Fluent was employed for handling the changing mesh through time. Nodes on the interface boundary were displaced with
\begin{equation}
\Delta \textbf{x}_i = \Delta t \textbf{v}_i
\end{equation}
where $\textbf{x}$ is the position vector composed of $x$ the horizontal and $y$ the vertical coordinates; using user-defined functions.

A node shuffle algorithm \citep{Hewett2017} was used to uniformly distribute the nodes around the interface at each mesh update. Without this algorithm, mesh quality degrades as the profile of the boundary morphs into its new shape and the simulation diverges.

The interior nodes were updated with a linearly elastic solid model (with the mesh smoothing based option in Fluent). Mesh motion was governed by
\begin{equation}
\nabla \cdot \underline{\boldsymbol{\sigma}} (\textbf{m}) = \underline{\textbf{0}}
\end{equation}
where $\underline{\boldsymbol{\sigma}}$ is the stress tensor and $\textbf{m}$ the mesh displacement vector,
\begin{equation}
\underline{\boldsymbol{\sigma}}(\textbf{m}) = E \text{tr}( \underline{\boldsymbol{\epsilon}}(\textbf{m})) \underline{\textbf{I}} + 2 G \underline{\boldsymbol{\epsilon}} (\textbf{m})
\end{equation}
where $\underline{\boldsymbol{\epsilon}}$ is the strain tensor,
\begin{equation}
\underline{\boldsymbol{\epsilon}} (\textbf{m}) = \frac{1}{2} (\nabla \textbf{m} + (\nabla \textbf{m})^\text{T})
\end{equation}
and was parameterised with Poisson's ratio
\begin{equation}\label{eqn:poisson}
\text{Po} = \frac{1}{2(1+G/E)}
\end{equation}
where $G$ is the shear modulus and $E$ the Young's modulus; where we used the default of $\text{Po} = \num{0.45}$ which gave robust mesh deformations and retained similar thermal boundary layer resolutions throughout the simulations.

\section{Results}
We first validated our model by using a phase change material with uniform density (no natural convection; conduction only) and compared the results with the analytical solution. Next, we simulated one case below the temperature threshold $T_w = \SI{8.0}{\celsius}$ where an inverted pear-shape interface developed, and one case above where an upright pear-shape melting front formed.

\subsection{Validation with uniform density}\label{sec:SectionA}
The Stefan problem is simplified when a uniform fluid density is employed such that no natural convection exists and the heat is transferred exclusively by conduction. Equation~\ref{eqn:conduction} describes the temperature field throughout the melting evolution for this reduced one-dimensional case. The temperature gradient in Equation~\ref{eqn:velocity} was calculated by numerically differentiating the temperature at the interface boundary.

\begin{figure}[t]\centering
  \includegraphics{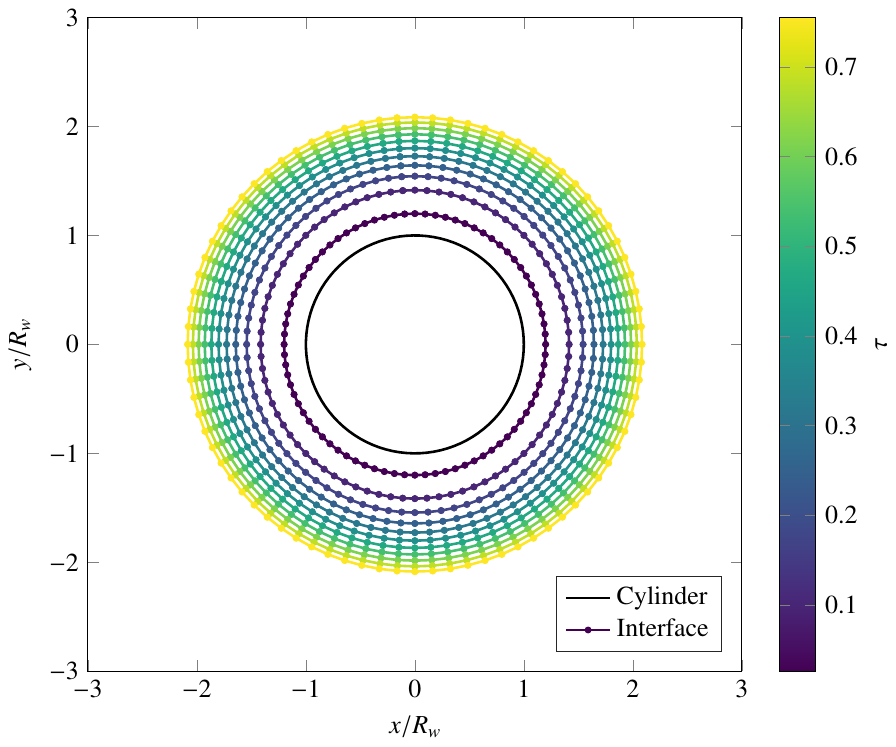}
  \caption{Melting interface evolution assuming uniform density with $T_w=\SI{2.3}{\celsius}$. Profiles are equally spaced in time.\label{fig:FigureC}}
\end{figure}

Evolution of the melting interface between the solid and liquid water phases are shown in Figure~\ref{fig:FigureC}. All time interface profiles are concentric with the cylinder, caused by uniform temperature gradients at the boundary, because no flow was generated within the fluid. The melting front is quickest during the initial period where the ice is almost in contact with the heated cylinder and slows as the interface recedes outward.

\begin{figure}[t]\centering
  \includegraphics{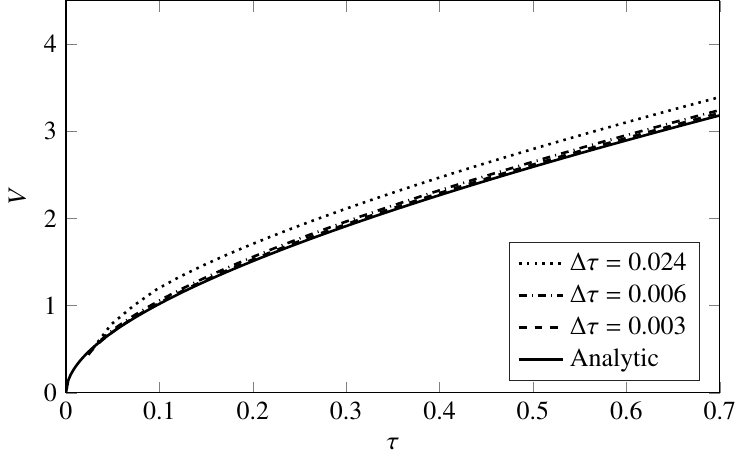}
  \caption{Time step convergence of melt volume over time for the uniform density case at $T_w = \SI{2.3}{\celsius}$.\label{fig:FigureA}}
\end{figure}

The rate of melt volume (Figure~\ref{fig:FigureA}) is greatest at the beginning and reduces over time; monotonically increasing. Simulations overestimate the melting rate because the temperature gradient was explicitly calculated at each step (essentially an Euler method). The coarse mesh time step of $\Delta \tau = 0.024$ significantly overestimates the melt volume and time steps of $\Delta \tau \leqslant 0.006$ agree well with the analytical solution.

Mesh independence was studied on a case by case basis as the flow features requiring varying mesh resolution levels differed across the cylinder temperatures and whether natural convection occurred or not. For example, the pure conduction case achieved mesh independence with a coarser mesh than the cases where natural convection occurred; recirculation of the flow needed to be resolved. Similarly, the mesh time step was converged for each case.

\subsection[Inverted pear-shape]{Inverted pear-shape ($T < \SI{8.0}{\celsius}$)}
\begin{figure}[t]\centering
  \includegraphics{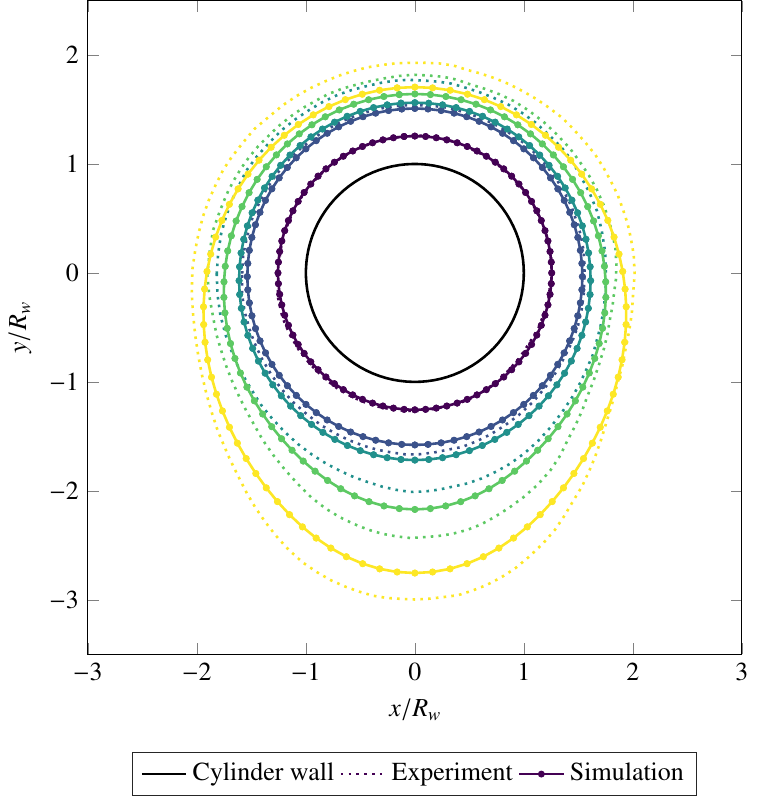}
  \caption{Melting interface profiles for $T_w = \SI{2.3}{\celsius}$, comparing simulation with experiment using paired dimensionless times of $\tau = 0.039,~0.172,~0.223,~0.350~\text{and}~0.524$ (extending outward from the cylinder respectively).\label{fig:FigureB}}
\end{figure}

A low cylinder temperature of $T_w = \SI{2.3}{\celsius}$ ($\text{Ste} = 0.029$, $\text{Ra} = 6700$) was chosen such that the coldest water had the lowest density (below the density inversion point). The melting interface, shown in Figure~\ref{fig:FigureB}, initially advances concentrically from the cylinder at the same rate as the uniform density case (Figure~\ref{fig:FigureC}).

\begin{figure}[t]\centering
  \includegraphics{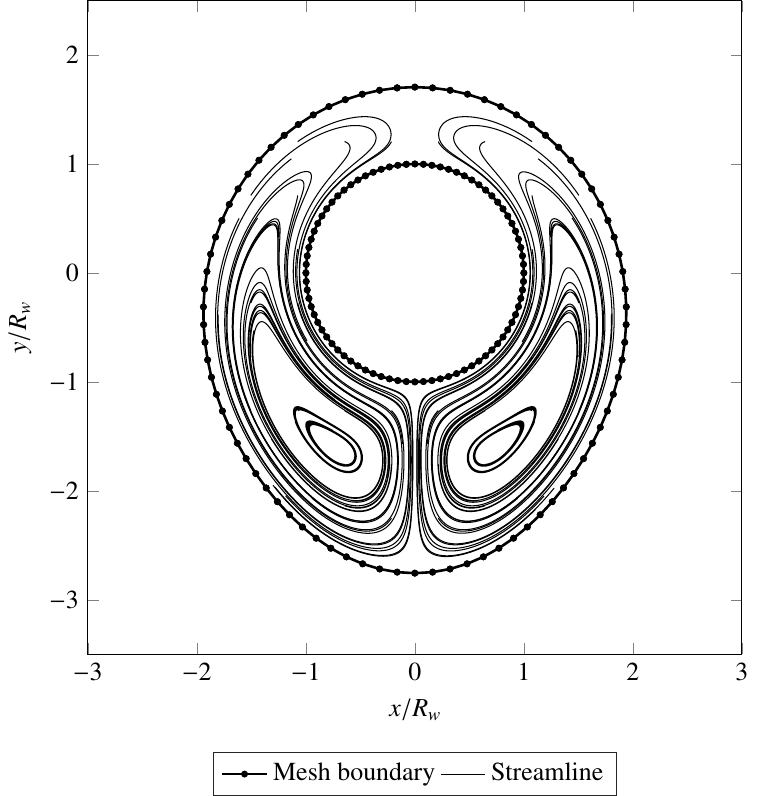}
  \caption{Streamlines for $T_w = \SI{2.3}{\celsius}$ at $\tau = 0.524$.\label{fig:FigureI}}
\end{figure}

Buoyancy driven flow develops as the melt volume increases causing recirculation as shown in Figure~\ref{fig:FigureI}. The warm water sinks (due to the density variation specified via Equation~\ref{eqn:density}), enhancing the melting rate at the base causing an inverted pear-shape form.

\begin{figure}[t]\centering
  \includegraphics{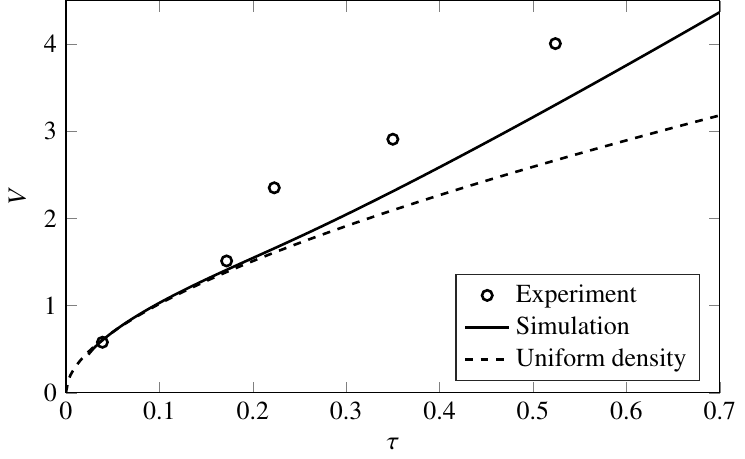}
  \caption{Molten ice volume over time with $T_w = \SI{2.3}{\celsius}$.\label{fig:FigureD}}
\end{figure}

Molten ice volume over time is shown in Figure~\ref{fig:FigureD} where both our simulation and the experiment \citep{White1986} closely follows the uniform density during the early stages of melting. This uniform density approximation begins to deviate at $\tau \approx 0.2$ where the fluid heat starts to be transported by natural convection in addition to conduction, causing the approximation to underestimate $V$.

A spike in $V$ at $\tau = 0.223$ was observed in the experiment (Figure~\ref{fig:FigureD}) which was not featured in our results. However, the slope of $V$ with $\tau$ match closely between the final two experiment data points and the second half of our simulation. Furthermore, the shape $R_i(\theta)$ closely resembles the experiment but lags slightly behind in time.

\subsection[Upright pear-shape]{Upright pear-shape ($T > \SI{8.0}{\celsius}$)}
\begin{figure}[t]\centering
  \includegraphics{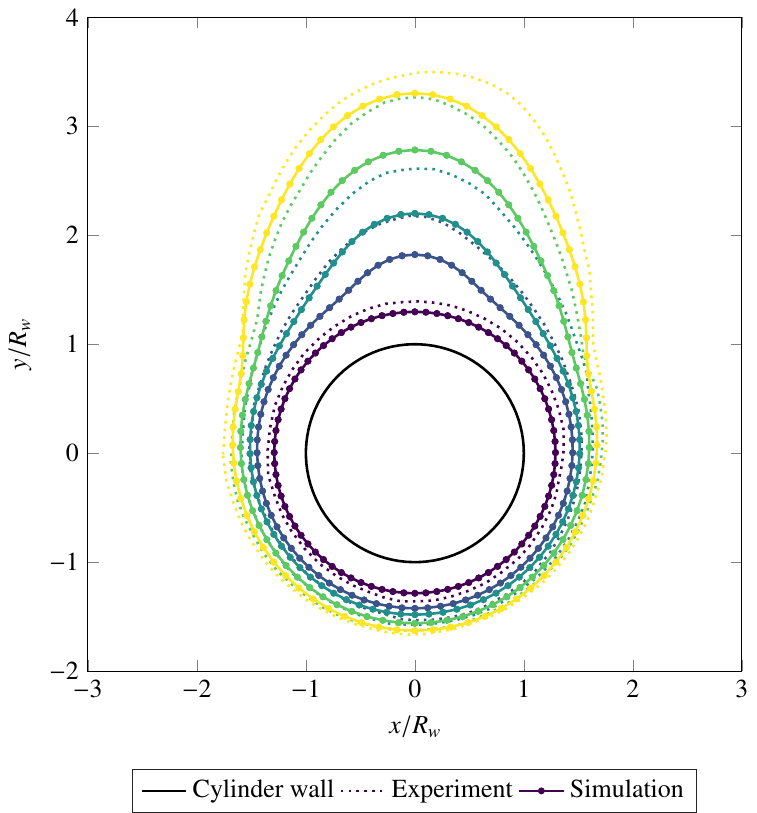}
  \caption{Melting interface profiles for $T_w = \SI{14.1}{\celsius}$, comparing simulation with experiment using paired dimensionless times of $\tau = 0.045,~0.122,~0.177,~0.272~\text{and}~0.370$ (extending outward from the cylinder respectively).\label{fig:FigureF}}
\end{figure}

The final case included the density inversion effects of water by prescribing a cylinder boundary temperature of $T_w = \SI{14.1}{\celsius}$ ($\text{Ste} = 0.178$, $\text{Ra} = 7400$). The melting interface for this case also begins with a concentrically evolving profile as shown in Figure~\ref{fig:FigureF}.

\begin{figure}[t]\centering
  \includegraphics{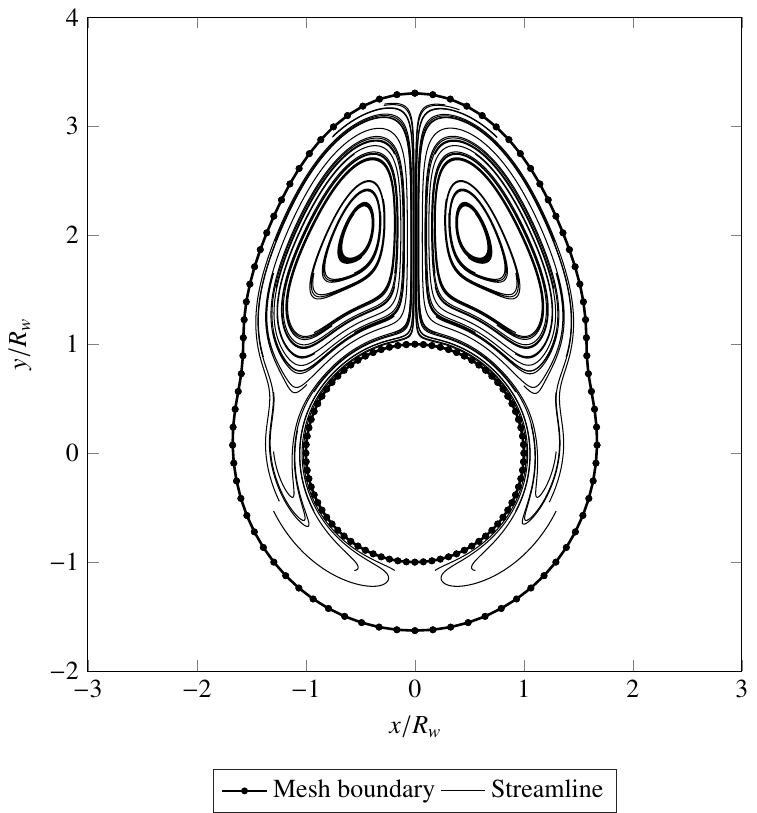}
  \caption{Streamlines for $T_w = \SI{14.1}{\celsius}$ at $\tau = 0.370$.\label{fig:FigureJ}}
\end{figure}

A thermal plume develops as the molten ice volume increases and this plume generates two counter-rotating vortices as shown in Figure~\ref{fig:FigureJ}; these vortices were also observed in experiments \citep{White1986}. This rising warm water and corresponding vortices causes an upright pear-shape profile to emerge.

\begin{figure}[t]\centering
  \includegraphics{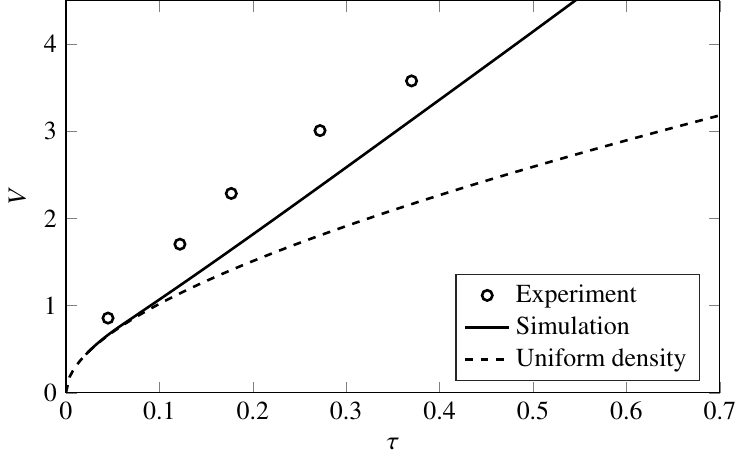}
  \caption{Molten ice volume over time with $T_w = \SI{14.1}{\celsius}$.\label{fig:FigureE}}
\end{figure}

The molten ice volume deviates from the uniform density approximation earlier for this warmer case as shown in Figure~\ref{fig:FigureE}. The ice melting rate observed in the experiments appear to exceed that of by pure conduction even in the early stages of the process at $\tau = 0.045$ whereas our simulations follow this rate until $\tau \approx 0.1$. Similar to the low temperature case, the slope of $V$ with $\tau$ match reasonably well between the experiment and our simulations during the natural convection dominated heat transfer regime.

Instantaneous interface profiles match closely with the experiment (Figure~\ref{fig:FigureF}) but are out of sync; slightly lagging in time. Small asymmetric features were observed in the experiment during the final time snapshots whereas our simulations predicted symmetric profiles about the vertical plane.

\section{Discussion}
The solid-liquid interface advanced radially outward from the heated cylinder in all cases; causing a monotonically increasing melt volume. Resolidification of the molten ice was not allowed in our model as a positive temperature gradient was present at the interface boundary (since $T > T_f$ throughout the domain); and resolidification was not observed in experiments \citep{White1986}. The conduction only case exhibited uniform melting rates as a function of $\theta$, resulting in concentric interface profiles, whereas the other two cases involving natural convection featured eccentricity in their interface shapes beyond the initial conduction dominated period. The location of eccentricity was determined by the opposite direction of the thermal plume: either above or below the cylinder.

\begin{figure}[t]\centering
  \includegraphics{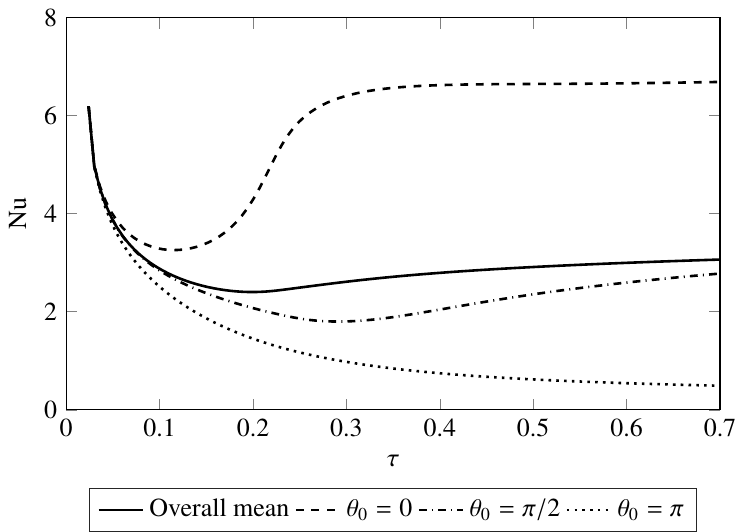}
  \caption{Variation of the local and total Nusselt numbers over time for the inverted pear-shape case ($T_w = \SI{2.3}{\celsius}$). Angle $\theta$ is measured from the base of the cylinder and is positive in the anti-clockwise direction.\label{fig:FigureG}}
\end{figure}

The local Nusselt number around the melting interface was defined as
\begin{equation}\label{eqn:Nusselt}
\text{Nu}_i = \frac{R_i}{T_f - T_w} \frac{dT}{dn} \big|_i
\end{equation}
where an area averaged radius of $R_i = \sqrt{A_i/\pi}$ was used to account for the time-dependent expansion of the domain. Figure~\ref{fig:FigureG} shows that the total heat transfer rapidly decreases in the early stages as the conduction heat transport mechanism accelerates the melting process. Heat transfer was uniform around the cylinder up until $\tau \approx 0.05$ where the local angle dependent $\text{Nu}$ numbers split and natural convection begins to develop. However, the melt volume rate deviates between the two cases at $\tau \approx 0.2$ (Figure~\ref{fig:FigureD}) indicating that the average melting rate is similar, regardless of the non-uniform melting rate, for $0.05 < \tau < 0.2$. Average $\text{Nu}$ appears to plateau toward the end of the simulation ($\tau > 0.6$), indicating that an equilibrium of the melting rate has been reached; also shown as the slope of $V$ with $\tau$ in Figure~\ref{fig:FigureD}.

Heat transfer by conduction dominated the early stages of the process and natural convection became a significant factor as the molten ice volume expanded, allowing recirculation to occur (Figures~\ref{fig:FigureI} and \ref{fig:FigureJ}). Experiments observed a spike in the rate of melting at the transition between the conduction and natural convection dominated heat transfer regimes. Our simulations did not capture this counter-intuitive local spike, possibly due to our simplified assumptions including 2-D steady flow. Local transient fluid motion induced from the transition from conduction to natural convection (unsteady flow) or 3-D effects may have been responsible for this enhanced melting rate. \citet{White1986} observed 3-D vortex motion in the later stages of the melting process causing ripples around the interface along the cylinder axis.

Standard uncertainties in measurements (either explicitly or implicitly via perturbing the rig) and values used for the thermophysical properties are other possible sources for the quantitative discrepancy between simulations and experiment. Melting rates and flow features are highly sensitive to the temperature of the heated cylinder. However, it is important to note that the interface shapes closely match throughout the melting process; and only the time dependent interface evolution at the flow regime transition differs.

The Rayleigh numbers of the two pear-shape cases involving natural convection were similar: $\num{6700}$ (inverted) and $\num{7400}$ (upright). However, each case was either side of the density inversion point of water. The inverted pear-shape had a positive monotonic relationship of density to temperature yielding a relatively simple recirculation flow pattern on each side of the pear. In contrast, the upright pear-shape encompassed the density inversion point causing more complex flow features such as counter-rotating vortices above the cylinder separated by the thermal plume (Figure~\ref{fig:FigureJ}). An instability of the thermal plume was observed in the experiment \citep{White1986} at $\tau \approx 0.3$ as the melt layer grew. These instabilities, and any unsteady flow features, are absent in our steady state model which could explain the discrepancy between the melt rates from $\tau = 0.3$ for the $T_w = \SI{14.1}{\celsius}$ (Figure~\ref{fig:FigureE}).

We assumed the flow field was in steady state for each discrete flow domain update (mesh step). The streamlines shown in images produced with an interferometer during the experiments \citep{White1986} indicate a mostly symmetric and steady flow field. However, slight perturbations and unsteady behaviour could cause greater heat mixing leading to faster melting, particularly for the $T_w > \SI{8.0}{\celsius}$ case. For example, assuming 2-D steady flow significantly overestimates the skin friction in the wake of an eroding cylinder evolving to a different shape than that of simulating with 3-D unsteady flow \citep{Hewett2017}. Another step would be required for including these unsteady effects: the time-averaged wall temperature gradient must first be established before calculating $\textbf{v}_i$ or the mesh would be deformed based on instantaneous local transients.

\section{Conclusion}
Numerical simulations were used to model the classical Stefan problem around a heated horizontal cylinder near the density inversion point of water by modelling both conduction and natural convection heat transfer. The melting front was explicitly tracked with a dynamic mesh and a node shuffle algorithm was employed to retain mesh quality through significant mesh deformation.

Heat transfer by conduction was dominant for the early stages of melting and natural convection played an important role as the melt zone increased. A stable steady flow field was found for the case below the density inversion point whereas a more complex and less stable flow was simulated when including the density inversion point. Pear-shaped melting interfaces developed as a thermal plume from the heated cylinder interacted with the solid-liquid boundary.

This paper provides a validation for modelling Stefan problems by tracking the melting front interface using only the local temperature gradient and fluid properties. The constitutive relation, along with the tools for handling the mesh deformation, form a useful approach for simulating this melting boundary problem found in latent heat thermal energy storage systems. This approach can also be applied to other scenarios of moving boundary problems.

\bibliography{MeltingCylinder}

\begin{thebibliography}{15}
\expandafter\ifx\csname natexlab\endcsname\relax\def\natexlab#1{#1}\fi
\providecommand{\url}[1]{\texttt{#1}}
\providecommand{\href}[2]{#2}
\providecommand{\path}[1]{#1}
\providecommand{\DOIprefix}{doi:}
\providecommand{\ArXivprefix}{arXiv:}
\providecommand{\URLprefix}{URL: }
\providecommand{\Pubmedprefix}{pmid:}
\providecommand{\doi}[1]{\href{http://dx.doi.org/#1}{\path{#1}}}
\providecommand{\Pubmed}[1]{\href{pmid:#1}{\path{#1}}}
\providecommand{\bibinfo}[2]{#2}
\ifx\xfnm\undefined \def\xfnm[#1]{\unskip,\space#1}\fi
\bibitem[{Akeiber et~al.(2016)Akeiber, Nejat, Majid, Wahid, Jomehzadeh,
  Zeynali~Famileh, Calautit, Hughes and Zaki}]{Akeiber2016}
\bibinfo{author}{Akeiber\xfnm[ H.]}, \bibinfo{author}{Nejat\xfnm[ P.]},
  \bibinfo{author}{Majid\xfnm[ M.Z.A.]}, \bibinfo{author}{Wahid\xfnm[ M.A.]},
  \bibinfo{author}{Jomehzadeh\xfnm[ F.]},
  \bibinfo{author}{Zeynali~Famileh\xfnm[ I.]}, \bibinfo{author}{Calautit\xfnm[
  J.K.]}, \bibinfo{author}{Hughes\xfnm[ B.R.]}, \bibinfo{author}{Zaki\xfnm[
  S.A.]}.
\newblock \bibinfo{title}{A review on phase change material ({PCM}) for
  sustainable passive cooling in building envelopes}.
\newblock \bibinfo{journal}{Renewable and Sustainable Energy Reviews}
  \bibinfo{year}{2016};\bibinfo{volume}{60}:\bibinfo{pages}{1470--1497}.
\bibitem[{Bathelt and Viskanta(1980)}]{Bathelt1980}
\bibinfo{author}{Bathelt\xfnm[ A.G.]}, \bibinfo{author}{Viskanta\xfnm[ R.]}.
\newblock \bibinfo{title}{Heat transfer at the solid-liquid interface during
  melting from a horizontal cylinder}.
\newblock \bibinfo{journal}{International Journal of Heat and Mass Transfer}
  \bibinfo{year}{1980};\bibinfo{volume}{23}(\bibinfo{number}{11}):\bibinfo{pages}{1493--1503}.
\bibitem[{Bathelt et~al.(1979)Bathelt, Viskanta and Leidenfrost}]{Bathelt1979}
\bibinfo{author}{Bathelt\xfnm[ A.G.]}, \bibinfo{author}{Viskanta\xfnm[ R.]},
  \bibinfo{author}{Leidenfrost\xfnm[ W.]}.
\newblock \bibinfo{title}{An experimental investigation of natural convection
  in the melted region around a heated horizontal cylinder}.
\newblock \bibinfo{journal}{Journal of Fluid Mechanics}
  \bibinfo{year}{1979};\bibinfo{volume}{90}(\bibinfo{number}{2}):\bibinfo{pages}{227--239}.
\bibitem[{Darzi et~al.(2012)Darzi, Farhadi and Sedighi}]{Darzi2012}
\bibinfo{author}{Darzi\xfnm[ A.R.]}, \bibinfo{author}{Farhadi\xfnm[ M.]},
  \bibinfo{author}{Sedighi\xfnm[ K.]}.
\newblock \bibinfo{title}{Numerical study of melting inside concentric and
  eccentric horizontal annulus}.
\newblock \bibinfo{journal}{Applied Mathematical Modelling}
  \bibinfo{year}{2012};\bibinfo{volume}{36}(\bibinfo{number}{9}):\bibinfo{pages}{4080--4086}.
\bibitem[{Farid et~al.(2004)Farid, Khudhair, Razack and Al-Hallaj}]{Farid2004}
\bibinfo{author}{Farid\xfnm[ M.M.]}, \bibinfo{author}{Khudhair\xfnm[ A.M.]},
  \bibinfo{author}{Razack\xfnm[ S.A.K.]}, \bibinfo{author}{Al-Hallaj\xfnm[
  S.]}.
\newblock \bibinfo{title}{A review on phase change energy storage: materials
  and applications}.
\newblock \bibinfo{journal}{Energy Conversion and Management}
  \bibinfo{year}{2004};\bibinfo{volume}{45}(\bibinfo{number}{9-10}):\bibinfo{pages}{1597--1615}.
\bibitem[{Gebhart and Mollendorf(1977)}]{Gebhart1977}
\bibinfo{author}{Gebhart\xfnm[ B.]}, \bibinfo{author}{Mollendorf\xfnm[ J.C.]}.
\newblock \bibinfo{title}{A new density relation for pure and saline water}.
\newblock \bibinfo{journal}{Deep Sea Research}
  \bibinfo{year}{1977};\bibinfo{volume}{24}(\bibinfo{number}{9}):\bibinfo{pages}{831--848}.
\newblock \DOIprefix\doi{http://dx.doi.org/10.1016/0146-6291(77)90475-1}.
\bibitem[{Herrmann et~al.(1984)Herrmann, Leidenfrost and
  Viskanta}]{Herrmann1984}
\bibinfo{author}{Herrmann\xfnm[ J.]}, \bibinfo{author}{Leidenfrost\xfnm[ W.]},
  \bibinfo{author}{Viskanta\xfnm[ R.]}.
\newblock \bibinfo{title}{Melting of ice around a horizontal isothermal
  cylindrical heat source}.
\newblock \bibinfo{journal}{Chemical Engineering Communications}
  \bibinfo{year}{1984};\bibinfo{volume}{25}(\bibinfo{number}{1-6}):\bibinfo{pages}{63--78}.
\bibitem[{Hewett and Sellier(2017)}]{Hewett2017}
\bibinfo{author}{Hewett\xfnm[ J.N.]}, \bibinfo{author}{Sellier\xfnm[ M.]}.
\newblock \bibinfo{title}{Evolution of an eroding cylinder in single and
  lattice arrangements}.
\newblock \bibinfo{journal}{Journal of Fluids and Structures}
  \bibinfo{year}{2017};\bibinfo{volume}{70}:\bibinfo{pages}{295--313}.
\newblock
  \DOIprefix\doi{http://dx.doi.org/10.1016/j.jfluidstructs.2017.01.011}.
\bibitem[{Ho and Chen(1986)}]{Ho1986}
\bibinfo{author}{Ho\xfnm[ C.J.]}, \bibinfo{author}{Chen\xfnm[ S.]}.
\newblock \bibinfo{title}{Numerical simulation of melting of ice around a
  horizontal cylinder}.
\newblock \bibinfo{journal}{International Journal of Heat and Mass Transfer}
  \bibinfo{year}{1986};\bibinfo{volume}{29}(\bibinfo{number}{9}):\bibinfo{pages}{1359--1369}.
\bibitem[{Moore(2017)}]{Moore2017}
\bibinfo{author}{Moore\xfnm[ M.N.J.]}.
\newblock \bibinfo{title}{Riemann-{H}ilbert problems for the shapes formed by
  bodies dissolving, melting, and eroding in fluid flows}.
\newblock \bibinfo{journal}{Communications on Pure and Applied Mathematics}
  \bibinfo{year}{2017};.
\bibitem[{Rieger et~al.(1982)Rieger, Projahn and Beer}]{Rieger1982}
\bibinfo{author}{Rieger\xfnm[ H.]}, \bibinfo{author}{Projahn\xfnm[ U.]},
  \bibinfo{author}{Beer\xfnm[ H.]}.
\newblock \bibinfo{title}{Analysis of the heat transport mechanisms during
  melting around a horizontal circular cylinder}.
\newblock \bibinfo{journal}{International Journal of Heat and Mass Transfer}
  \bibinfo{year}{1982};\bibinfo{volume}{25}(\bibinfo{number}{1}):\bibinfo{pages}{137--147}.
\bibitem[{Sparrow et~al.(1978)Sparrow, Schmidt and Ramsey}]{Sparrow1978}
\bibinfo{author}{Sparrow\xfnm[ E.M.]}, \bibinfo{author}{Schmidt\xfnm[ R.R.]},
  \bibinfo{author}{Ramsey\xfnm[ J.W.]}.
\newblock \bibinfo{title}{Experiments on the role of natural convection in the
  melting of solids}.
\newblock \bibinfo{journal}{Journal of Heat Transfer}
  \bibinfo{year}{1978};\bibinfo{volume}{100}(\bibinfo{number}{1}):\bibinfo{pages}{11--16}.
\bibitem[{Stefan(1891)}]{Stefan1891}
\bibinfo{author}{Stefan\xfnm[ J.]}.
\newblock \bibinfo{title}{{\"U}ber die theorie der eisbildung, insbesondere
  \"uber die eisbildung im polarmeere}.
\newblock \bibinfo{journal}{Annalen der Physik}
  \bibinfo{year}{1891};\bibinfo{volume}{278}(\bibinfo{number}{2}):\bibinfo{pages}{269--286}.
\bibitem[{White et~al.(1986)White, Viskanta and Leidenfrost}]{White1986}
\bibinfo{author}{White\xfnm[ D.]}, \bibinfo{author}{Viskanta\xfnm[ R.]},
  \bibinfo{author}{Leidenfrost\xfnm[ W.]}.
\newblock \bibinfo{title}{Heat transfer during the melting of ice around a
  horizontal, isothermal cylinder}.
\newblock \bibinfo{journal}{Experiments in Fluids}
  \bibinfo{year}{1986};\bibinfo{volume}{4}(\bibinfo{number}{3}):\bibinfo{pages}{171--179}.
\bibitem[{White et~al.(1977)White, Bathelt, Leidenfrost and
  Viskanta}]{White1977}
\bibinfo{author}{White\xfnm[ R.D.]}, \bibinfo{author}{Bathelt\xfnm[ A.G.]},
  \bibinfo{author}{Leidenfrost\xfnm[ W.]}, \bibinfo{author}{Viskanta\xfnm[
  R.]}.
\newblock \bibinfo{title}{Study of heat transfer and melting front from a
  cylinder imbedded in a phase change material}.
\newblock \bibinfo{journal}{American Society of Mechanical Engineers (Paper)}
  \bibinfo{year}{1977};\bibinfo{volume}{77}.

\end{thebibliography}

\end{document}